\documentclass[referee]{raa}            

\usepackage{graphicx,times}             
\usepackage{natbib}
\bibpunct{(}{)}{;}{a}{}{,}

\begin{document}

   \title{Seismic study of solar convection and overshooting: \\results of nonlocal convection}

   \volnopage{Vol.0 (200x) No.0, 000--000}      
   \setcounter{page}{1}          

   \author{Chunguang Zhang
      \inst{1,2}
   \and Licai Deng
      \inst{1}
   \and Darun Xiong
      \inst{3}
   \and J{\o}rgen Christensen-Dalsgaard
      \inst{4}
   }

   \institute{Key Laboratory of Optical Astronomy, National Astronomical Observatories, Chinese Academy of Sciences,
              Beijing 100012, China; {\it cgzhang@nao.cas.cn}\\
        \and
              University of Chinese Academy of Sciences, Beijing 100049, China\\
        \and
              Purple Mountain Observatory, Chinese Academy of Sciences, Nanjing 210008, China\\
        \and
              Stellar Astrophysics Centre, Department of Physics and Astronomy, Aarhus University, Ny Munkegade 120,
              DK-8000 Aarhus C, Denmark
             }

   \date{Received~~2013 month day; accepted~~2013~~month day}

\abstract{Local mixing-length theory is incapable of describing nonlocal phenomena in stellar convection, such as overshooting. Therefore standard solar models constructed with the local mixing-length theory deviate significantly from the Sun at the boundaries of the convection zone, where convection becomes less efficient and nonlocal effects are important. The differences between observed and computed frequencies come mainly from the near-surface region, while the localized sound-speed difference is just below the convective envelope. In this paper we compute a solar envelope model using Xiong's nonlocal convection theory, and carry out helioseismic analysis. The nonlocal model has a smooth transition at the base of the convection zone, as revealed by helioseismology. It reproduces solar frequencies more accurately, and reduces the localized sound-speed difference between the Sun and standard solar models.
}

   \authorrunning{C. Zhang, L. Deng, D. Xiong \& J. Christensen-Dalsgaard}            
   \titlerunning{Seismic study of solar convection and overshooting}                   

   \maketitle

%
%
\section{INTRODUCTION}           
\label{sect:intro}
Stellar convection is generally expected to penetrate beyond the convectively unstable boundaries defined by the Schwarzschild criterion. This phenomenon, known as overshooting, has been observed at the surface of the Sun. Overshooting at the bottom of a convective envelope or from a convective core can have crucial influences on the structure and evolution of stars. In the Sun, it is thought to be closely related to the solar tachocline and solar dynamo. Although a lot of effort has been made to model the solar overshooting zone, there are still major disagreements among these models regarding the extent and importance of overshooting.

\newpage
In stellar modeling, the mixing-length theory \citep[MLT;][]{BV1958} is usually used to describe convection. It works quite well deep inside the convection zone because the convective energy transport is very efficient there, and the temperature stratification is nearly adiabatic regardless of the details of the convection theory. However, its local character makes MLT inadequate for the treatment of overshooting. In MLT, convective motions stop suddenly at the boundaries of the convection zone; the nonlocal diffusion of convection, the turbulent pressure, and the turbulent kinetic energy are ignored. Therefore the standard solar model (SSM) constructed with MLT shows clear differences from the Sun at both boundaries of the convection zone.

Convection becomes inefficient near the upper boundary of the solar convection zone, and the temperature gradient $\nabla$ deviates significantly from its adiabatic value $\nabla_\mathrm{ad}$. The nonlocal transport of convective energy and momentum has important effects on the thermal and dynamic structure of this near-surface region, but MLT fails to predict due to its local approximation. Helioseismology has shown that the frequency differences between the Sun and SSM arise predominantly from improper modeling of the near-surface layers of the Sun \citep{ModelS}. Similar systematic offsets have been found for solar-like stars. Therefore a more realistic description of convection is required in order to improve the modeling of the near-surface region and match the observed frequencies.

At the lower boundary of the solar convection zone, MLT predicts a sudden switch from the adiabatic temperature gradient $\nabla_\mathrm{ad}$ to the radiative temperature gradient $\nabla_\mathrm{rad}$. Because convection stops abruptly at the boundary, gravitational settling of helium and heavy elements causes a steep composition gradient, which leads to a localized sharp feature in the sound-speed difference between the Sun and SSM, as revealed by helioseismic inversions \citep{Basu1997}. Overshooting would cause chemical mixing outside the convection zone and thereby remove the composition gradient, but it cannot be incorporated within the framework of local MLT.

To solve the problem of modeling overshooting, a series of nonlocal MLT have been developed \citep{SS1973, Maeder1975, BCB1981, Langer1986}. However, their conclusions differ widely. Some models predicted extensive overshooting \citep{SS1973, Maeder1975, BCB1981}, while others found overshooting to be negligibly small \citep{SS1965, Langer1986}. This was caused by the different built-in assumptions. As pointed out by \citet{Renzini1987}, all these theories of the ballistic type had physical inconsistencies introduced by a confusion between local and nonlocal quantities; therefore, their results on overshooting were preordained instead of calculated within the theory. Moreover, the simple overshooting models constructed with nonlocal MLT have been disproved by helioseismic investigations because they cannot give the smooth stratification as shown to be the case in the Sun \citep{MCDT1994, BAN1994, RV1994}.

\newpage
Another theoretical approach toward solving this problem is the Reynolds Stress Model \citep[RSM;][]{Xiong1979, Xiong1981, Xiong1989, XCD1997, Canuto1993, CD1997, CD1998, LY2007}. It starts from the basic hydrodynamic equations by dividing the physical variables, such as velocity, temperature, and density, into averaged values and turbulent fluctuations. As the Navier-Stokes equations are non-linear, Reynolds averaging yields an unclosed hierarchy of moment equations, and the third-order moments represent the nonlocal transport of turbulent energy and momentum. Therefore, to form a working nonlocal formulation, closure approximations must be made at the third or higher order. Different approximations lead to different nonlocal convection theories. Most of the approximations introduce free parameters. Like the $\alpha$ parameter in MLT, they have to be adjusted to meet observational constraints.

Theories of the RSM type, although usually complicated in applications, are more physically grounded than MLT, and are capable of dealing with nonlocal phenomena like overshooting. In the working equations of Xiong's theory, the moment equations of auto- and cross-correlations of turbulent velocity and temperature are closed at the third order with a gradient-type diffusion approximation, in which the third-order moments are expressed as the gradients of the corresponding second-order moments. It has two dimensionless parameters $c_1$ and $c_2$ related to turbulent dissipation and diffusion, respectively. Xiong's theory has been successfully used in modeling stellar structure, evolution, and oscillation \citep{Xiong1986, XC1992, XD2007, XD2009}.

In this paper, we use Xiong's theory to construct a solar model with updated input physics to study the structure and properties of solar convection and overshooting with helioseismic methods. To avoid extra complications such as nuclear reactions and chemical evolution in the core, our calculations have been limited to envelope models with the bottom at $r=0.3R$, where $r$ is the distance from the center and $R$ is the photospheric radius of the Sun. Section \ref{sect:ModelNL} presents the details of the envelope model and its differences from SSM. In Section \ref{sect:Helioseismology}, we calculate the oscillation frequencies of the nonlocal model and carry out the seismic analysis. A summary and discussion are presented in Section \ref{sect:Conclusion}.

\section{THE SOLAR ENVELOPE MODEL}
\label{sect:ModelNL}
\subsection{Model Construction}
\label{sect:model}
\citet{XD2001} computed a static solar model using the radiation-hydrodynamic equations \citep{Xiong1989}. They used this model to discuss the properties of solar convection and overshooting. To further study solar convection with helioseismology, we recompute a solar envelope model with updated input physics. The computation uses the OPAL equation of state \citep{OPALEOS} and opacities \citep{OPAL}. For temperatures lower than $6000$ K, opacity tables of \citet{Ferguson} are taken. Diffusion of helium and heavy elements is not included, thus our envelope model is chemically uniform. We use Model S \citep{ModelS} as the reference SSM. Our nonlocal model, in the following Model NL, has been calibrated to the same basic properties of Model S, including the photospheric radius $R = 6.9599 \times 10^{10} \rm{cm}$, surface luminosity $L = 3.846 \times 10^{33} \rm{erg\ s^{-1}}$, and a mass ratio $Z/X = 0.0245$ between heavy elements and hydrogen \citep{GN93}.

\newpage
In order to compare with Model S and the Sun, Model NL should have comparable depth of the convection zone. In the local description, the convectively unstable region can be defined according to many equivalent criteria, such as $\nabla > \nabla_\mathrm{ad}$, $\nabla_\mathrm{rad} > \nabla_\mathrm{ad}$, or $F_\mathrm{c} > 0$, where $F_\mathrm{c}$ is the convective (enthalpy) flux. However, in nonlocal theories, the boundaries defined by these criteria do not generally coincide \citep{Canuto1997, XD2001, BM2010}. \citet{DX2008} argued that a proper definition of the boundaries of the convection zone should be the place where $F_\mathrm{c}$ (or equivalently, the correlation of turbulent velocity and temperature) changes sign, so that the local and nonlocal models with the same depth of the convection zone would have similar structure, and the overshooting zone could be consistently defined as the region with negative $F_\mathrm{c}$. In the nonlocal theory, it is $F_\mathrm{c}=0$, instead of the Schwarzschild criterion, that defines the point of neutral buoyancy. Convective motions are driven by the buoyancy force in the convection zone ($F_\mathrm{c}>0$), and dissipate in the overshooting zone ($F_\mathrm{c}<0$). By this criterion, the lower boundary of the convection zone in Model NL has been calibrated to $0.7123R$, which is consistent with helioseismic results \citep{JCD1991,BA1997}.

\subsection{The Superadiabatic Layer}
\label{sect:SAL}
\begin{figure}[htbp]
\centering
\includegraphics[width=0.8\textwidth]{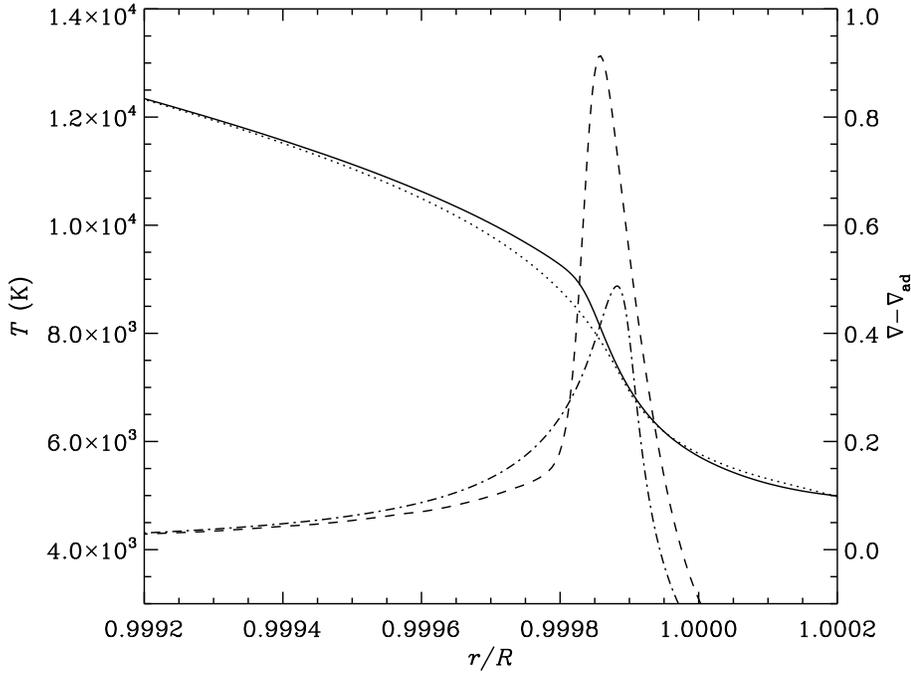}
\caption{Temperature $T$ (left ordinate) and superadiabaticity $\nabla-\nabla_\mathrm{ad}$ (right ordinate) vs. fractional radius $r/R$ in the SAL. Solid line: $T$ in Model NL; dotted line: $T$ in Model S; dashed line: $\nabla-\nabla_\mathrm{ad}$ in Model NL; and dash-dotted line: $\nabla-\nabla_\mathrm{ad}$ in Model S.}
\label{fig:sal_tn}
\end{figure}

Most of the solar convection zone is nearly adiabatic with $\nabla$ slightly greater than $\nabla_\mathrm{ad}$ due to the high efficacy of convection. However, the near-surface region is substantially superadiabatic, where $\nabla-\nabla_\mathrm{ad}$ is of the order unity. As a transition between the adiabatic region and the radiative atmosphere, this thin superadiabatic layer (SAL) is important in solar modeling. In one-dimensional (1-D) models, the overall structure of the convection zone depends on the integrated properties of the SAL \citep{GW1976}. The entropy jump across the SAL is closely related to the efficacy of convection, which in MLT is controlled by the mixing length. An increase of the mixing length renders convective transport more efficient. This means $\nabla-\nabla_\mathrm{ad}$ in the SAL becomes smaller, and therefore the entropy jump is reduced, which leads to a smaller radius of the model. Fig. \ref{fig:sal_tn} compares the temperature $T$ and the superadiabatic temperature gradient $\nabla-\nabla_\mathrm{ad}$ between Model NL and Model S. Model NL shows a steeper temperature gradient in the SAL, which is consistent with the empirical solar atmosphere \citep{GNKC1971} and numerical simulations \citep{Kupka2009, Beeck2012}. The large temperature gradient produces a high peak in $\nabla-\nabla_\mathrm{ad}$, indicating inefficient convection in terms of energy transport. MLT overestimates the efficacy of convection in the SAL, hence the peak in $\nabla-\nabla_\mathrm{ad}$ is lower in Model S.

\begin{figure}[htbp]
\centering
\includegraphics[width=0.8\textwidth]{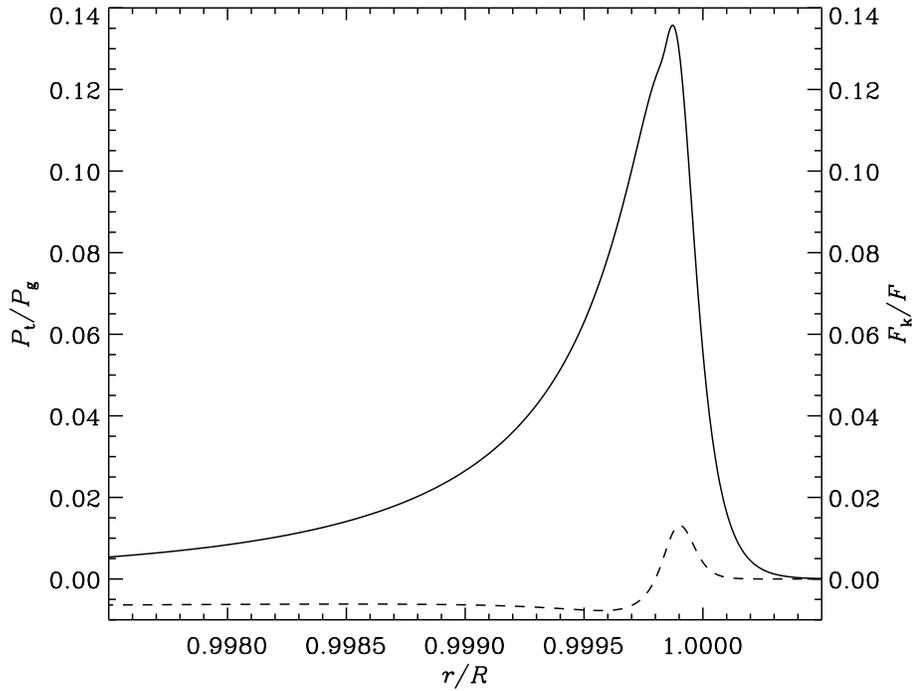}
\caption{Ratio of turbulent pressure to gas pressure $P_\mathrm{t}/P_\mathrm{g}$ (solid line, left ordinate) and fractional turbulent kinetic energy flux $F_\mathrm{k}/F$ (dashed line, right ordinate) vs. fractional radius $r/R$ in Model NL.}
\label{fig:sal_pf}
\end{figure}

The structure of the SAL is also affected by the dynamical effects of convection. However, in MLT the turbulent pressure $P_\mathrm{t}$ is neglected. Fig. \ref{fig:sal_pf} shows the ratio between turbulent pressure $P_\mathrm{t}$ and gas pressure $P_\mathrm{g}$ in the near-surface part of Model NL. $P_\mathrm{t}$ increases from almost zero to a maximum of $14$ percent near the peak of the SAL. Therefore it has significant contribution to the total pressure support against gravity.

\begin{figure}[htbp]
\centering
\includegraphics[width=0.8\textwidth]{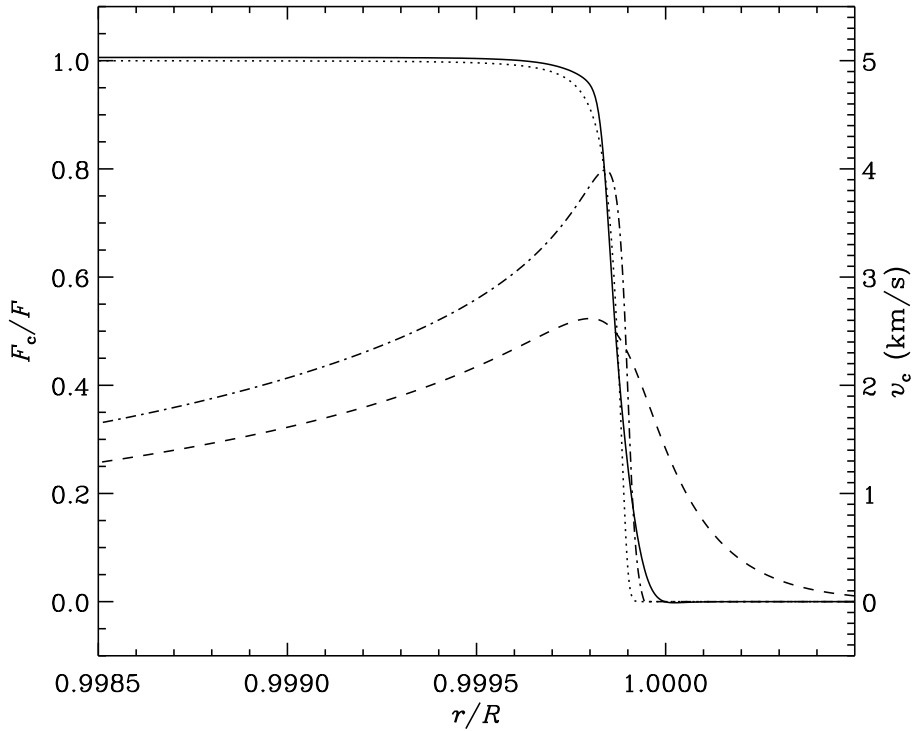}
\caption{Fractional convective flux $F_\mathrm{c}/F$ (left ordinate) and convective velocities $v_\mathrm{c}$ (right ordinate) vs. fractional radius $r/R$ in the SAL. Solid line: $F_\mathrm{c}/F$ in Model NL; dotted line: $F_\mathrm{c}/F$ in Model S; dashed line: $v_\mathrm{c}$ in Model NL; and dash-dotted line: $v_\mathrm{c}$ in Model S.}
\label{fig:sal_fv}
\end{figure}

\newpage
The turbulent kinetic energy flux $F_\mathrm{k}$, in units of the total energy flux $F$, is also plotted in Fig. \ref{fig:sal_pf}. $F_\mathrm{k}/F$ is less than $1.5\%$, thus its influence on the energy transport is very limited. However, a part of $F_\mathrm{k}$ runs into the overshooting region due to the nonlocal turbulent diffusion, therefore the convective energy flux $F_\mathrm{c}$ and the convective velocity $v_\mathrm{c}$ in the SAL are lower than MLT predicts, as shown in Fig. \ref{fig:sal_fv}. In model NL, velocity fluctuations occur outside the convection zone because of overshooting, while in Model S, $v_\mathrm{c}$ is set to zero above the boundary.

Solar near-surface convection has also been studied by means of 3-D numerical simulations \citep{Abbett1997, KC1998, SN1998, CO5BOLD}. Their results on the temperature profile, turbulent pressure, and overshooting are in good general agreement with ours as shown in Fig. \ref{fig:sal_tn} - \ref{fig:sal_fv}. However, some simulations predict very large $F_\mathrm{k}$ \citep{KC1998, SN1998}, usually a magnitude larger than we have shown here. Therefore $F_\mathrm{k}$ contributes significantly to the total energy transport in these models. Whether such a difference comes from the approximations adopted in the simulations or from the closure scheme in the RSM requires further investigation. We should also note that numerical simulations still show some distinct differences from each other \citep{Kupka2009, TBD2012} as the result of the different resolutions and approximations in use.

\newpage
\subsection{Solar Lower Overshooting Zone}
\label{sect:LOZ}
\begin{figure}[htbp]
\centering
\includegraphics[width=0.7\textwidth]{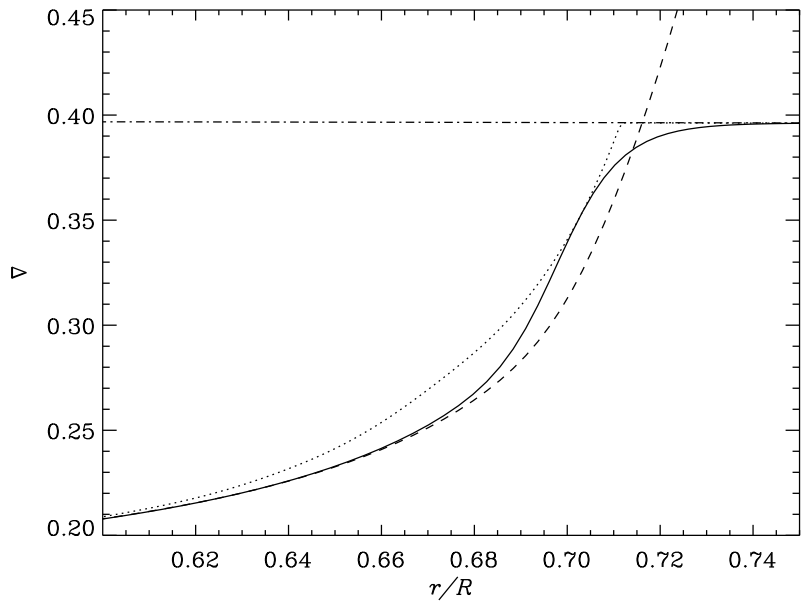}
\caption{Temperature gradients $\nabla$ (solid line), $\nabla_\mathrm{rad}$ (dashed line), and $\nabla_\mathrm{ad}$ (dash-dotted line) in Model NL. For comparison, the dotted line shows $\nabla$ in Model S.}
\label{fig:loz_n}
\end{figure}

\begin{figure}[htbp]
\centering
\includegraphics[width=0.7\textwidth]{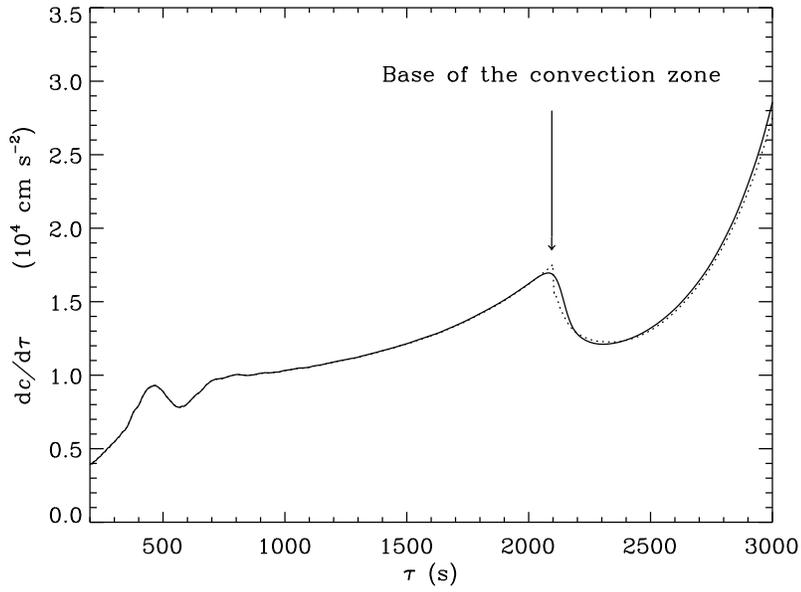}
\caption{Derivative of the sound speed vs. acoustic depth $\tau$ in Model NL (solid line) and Model S (dotted line).}
\label{fig:loz_cderiv}
\end{figure}

The transition at the bottom of the solar convection zone in MLT models is rather simple: $\nabla$ changes abruptly from $\nabla_\mathrm{ad}$ in the convective envelope to $\nabla_\mathrm{rad}$ in the radiative interior, and this switch defines the lower boundary of the convection zone. If overshooting is incorporated using nonlocal MLT with a nearly adiabatic extension from the convection zone, the abruptness of the transition becomes even more serious and causes discontinuity in the derivative of the sound speed $c$. However, in Model NL, $\nabla$ is already subadiabatic before reaching the boundary of the convection zone, thus the transition is smooth, as shown in Fig. \ref{fig:loz_n}. The lower overshooting zone (LOZ) is characterized by a subadiabatic but superradiative temperature gradient ($ \nabla_\mathrm{rad} < \nabla < \nabla_\mathrm{ad}$). As a result of the smooth stratification, the derivative of the sound speed in the LOZ in Model NL is continuous. This is illustrated in Fig. \ref{fig:loz_cderiv}, where the sound-speed derivative is plotted against the acoustic depth:
\begin{equation}
\tau=\int_{r}^{R}\frac{\mathrm{d}r}{c}.
\end{equation}

\begin{figure}[htbp]
\centering
\includegraphics[width=0.8\textwidth]{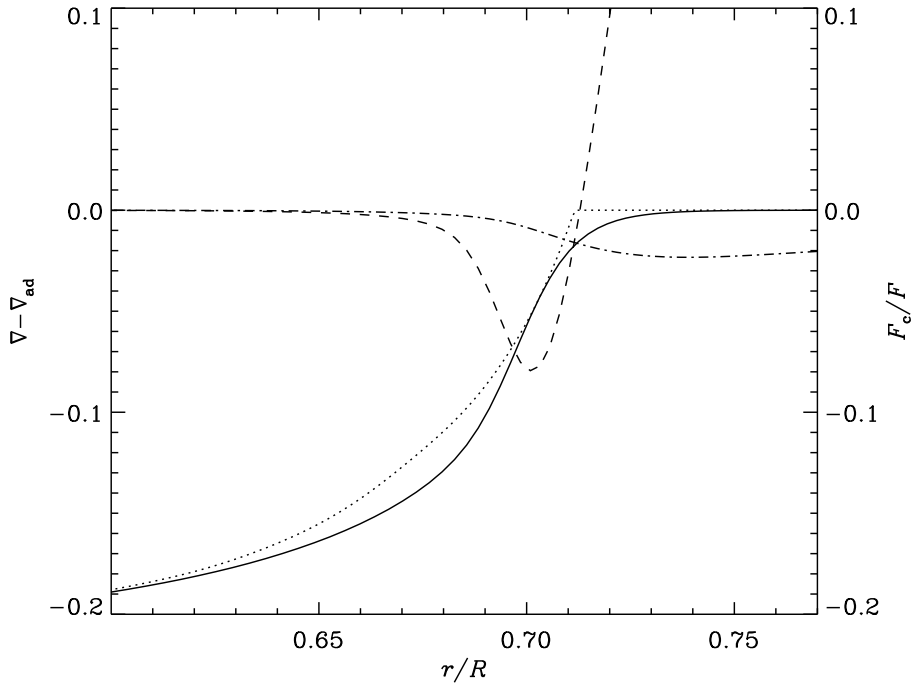}
\caption{Superadiabaticity and fractional energy fluxes vs. fractional radius $r/R$ in the LOZ. Left ordinate: superadiabatic temperature gradient $\nabla-\nabla_\mathrm{ad}$ in Model NL (solid line) and Model S (dotted line). Right ordinate: convective energy flux $F_\mathrm{c}$ (dashed line) and turbulent kinetic energy flux $F_\mathrm{k}$ (dash-dotted line) in Model NL, in units of the total flux $F$.}
\label{fig:loz_nf}
\end{figure}

Fig. \ref{fig:loz_nf} shows the superadiabatic temperature gradient and fractional energy fluxes in Model NL. As in the SAL, $F_\mathrm{k}$ is very small (about $2\%$ of the total energy flux). The convective boundary defined by $F_\mathrm{c}=0$ in Model NL is similar to that defined by the Schwarzschild criterion in Model S. In the LOZ, the convective energy flux $F_\mathrm{c}$ is negative, and the radiative energy flux $F_\mathrm{r}$ is larger than the total flux of the Sun. As a result, the temperature in the overshooting zone will increase. Therefore the sound speed in this region is higher than MLT predicts.

In the LOZ, turbulent velocity and temperature penetrate more deeply compared with the convective flux \citep{DX2008}, therefore convective mixing is very efficient. Overshooting extends the fully mixed region and may be the dominant mechanism of solar lithium depletion \citep{XD2009}.

\newpage
\section{Seismic Analysis}
\label{sect:Helioseismology}
\subsection{Frequencies of p-mode Oscillation}
\label{sect:fdiff}
Adiabatic oscillation frequencies are calculated for both models, using the Aarhus adiabatic oscillation package \citep{JCD2008}. Since Model NL is an envelope model, it cannot reproduce the deeply penetrating modes accurately. In our following analysis, only the modes with lower turning point $r_\mathrm{t} > 0.4R$ are considered. The observed frequencies are from the SOHO spacecraft \citep{SOHO}.

\begin{figure}[htbp]
\centering
\includegraphics[width=0.45\textwidth]{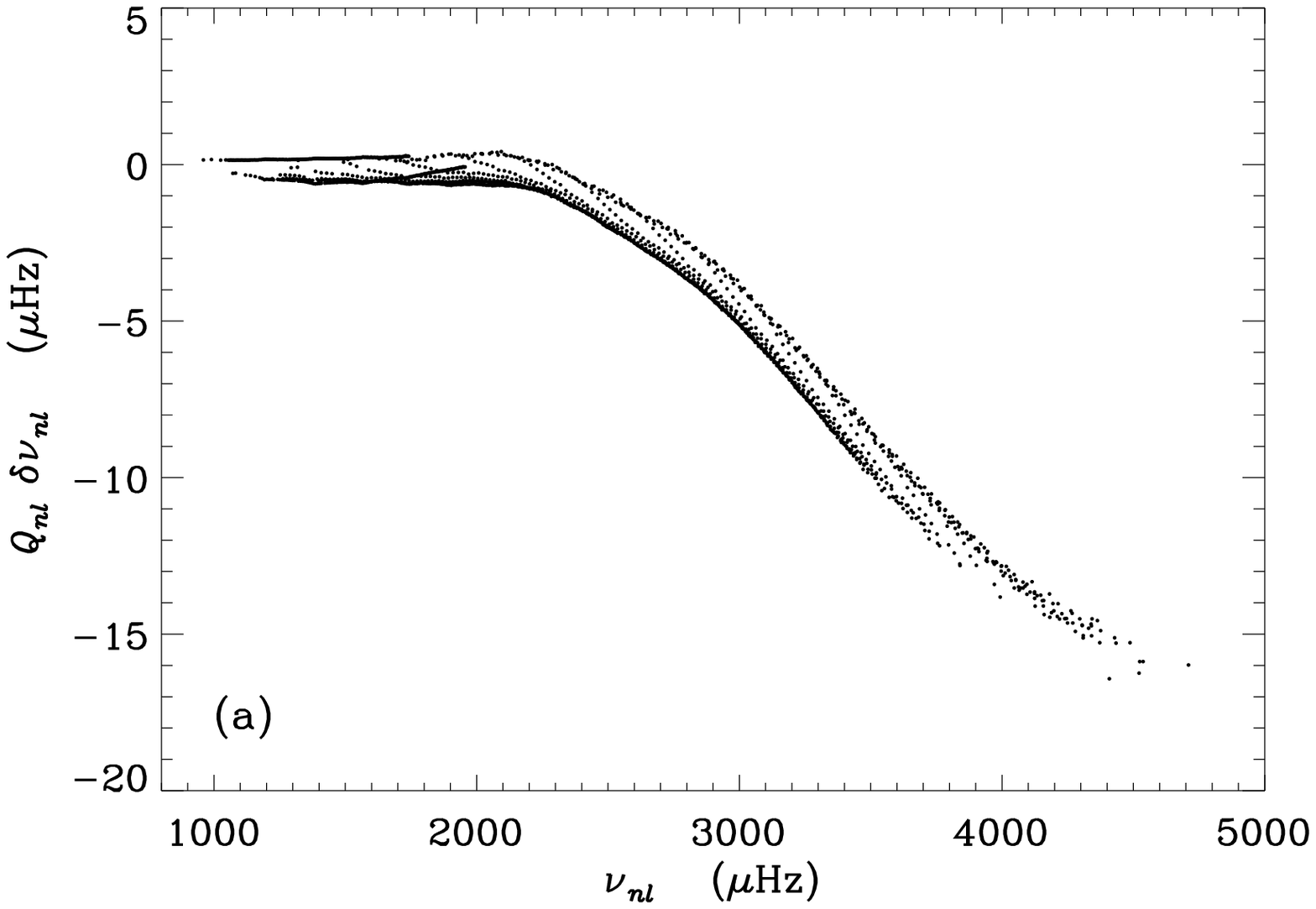}
\includegraphics[width=0.45\textwidth]{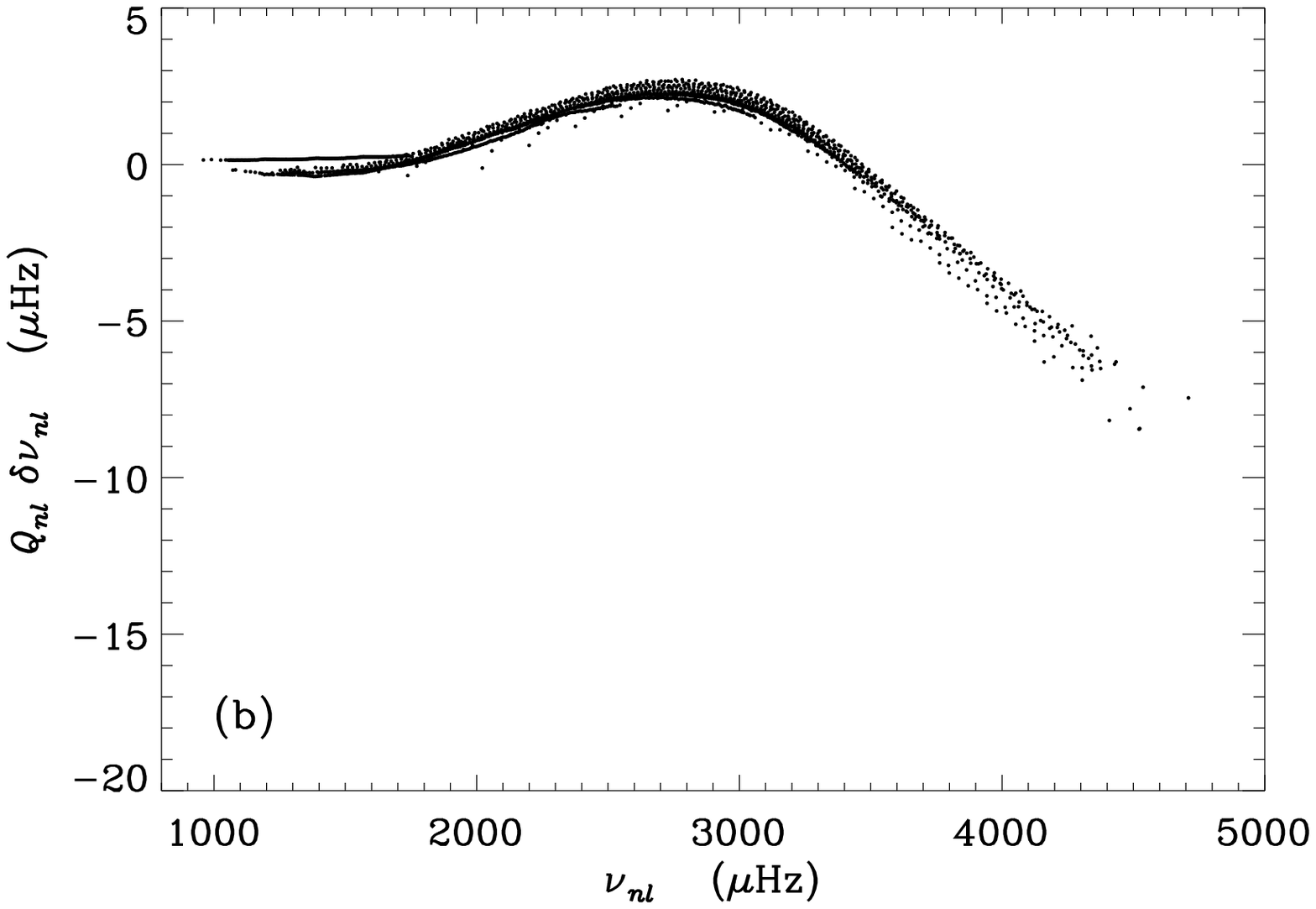}
\caption{Scaled frequency differences $Q_{nl}\delta \nu_{nl}$ between the Sun and models, in the sense (Sun) - (Model). Panel (a) shows $Q_{nl}\delta \nu_{nl}$ between the Sun and Model S, and panel (b) shows $Q_{nl}\delta \nu_{nl}$ between the Sun and Model NL.}
\label{fig:fdiff}
\end{figure}

\citet{ZDXCD2012} compared the frequencies between observation and the models. For the convenience of the following discussion, we reproduce in Fig.\ref{fig:fdiff} the scaled frequency differences $Q_{nl}\delta \nu_{nl}$, where $Q_{nl}$ is the inertia ratio \citep{ModelS} and $\delta\nu_{nl}$ is the difference between observed and computed frequency $\nu_{nl}$ with radial order $n$ and spherical degree $l$. For Model S, $Q_{nl}\delta \nu_{nl}$ is largely a function of frequency, which indicates that the differences between the Sun and Model S are dominated by the near-surface errors \citep{CDT1997}. However, Fig. \ref{fig:fdiff}(a) also shows two distinct branches, suggesting a depth dependence of the modes \citep{ModelS}, which is caused by the localized sound-speed difference as revealed by inversion.

In Fig. \ref{fig:fdiff}(b), $Q_{nl}\delta \nu_{nl}$ between the Sun and Model NL is much reduced compared to Fig. \ref{fig:fdiff}(a), especially at the high-frequency end, and the depth dependence is removed. The remaining $l$-independent differences in Fig. \ref{fig:fdiff}(b) suggest they are still closely related to the near-surface layers. Further improvement may come from including nonadiabatic effects in the calculation of model frequencies \citep{CX1997, Houdek2010, Grigahcene2012} because the interaction between convection and oscillations in the near-surface region also has important effects on the frequencies.

\begin{figure}[htbp]
\centering
\includegraphics[width=0.45\textwidth]{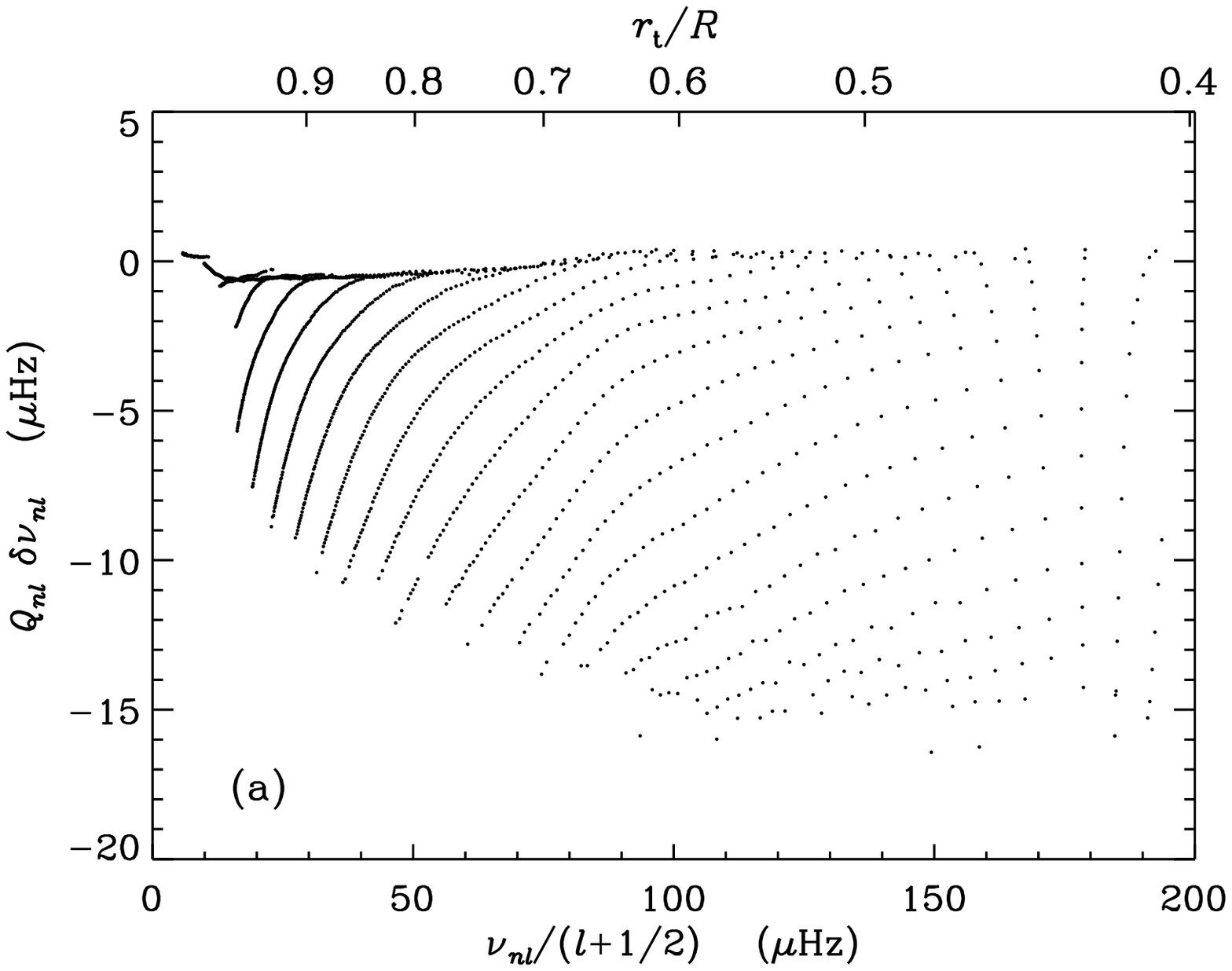}
\includegraphics[width=0.45\textwidth]{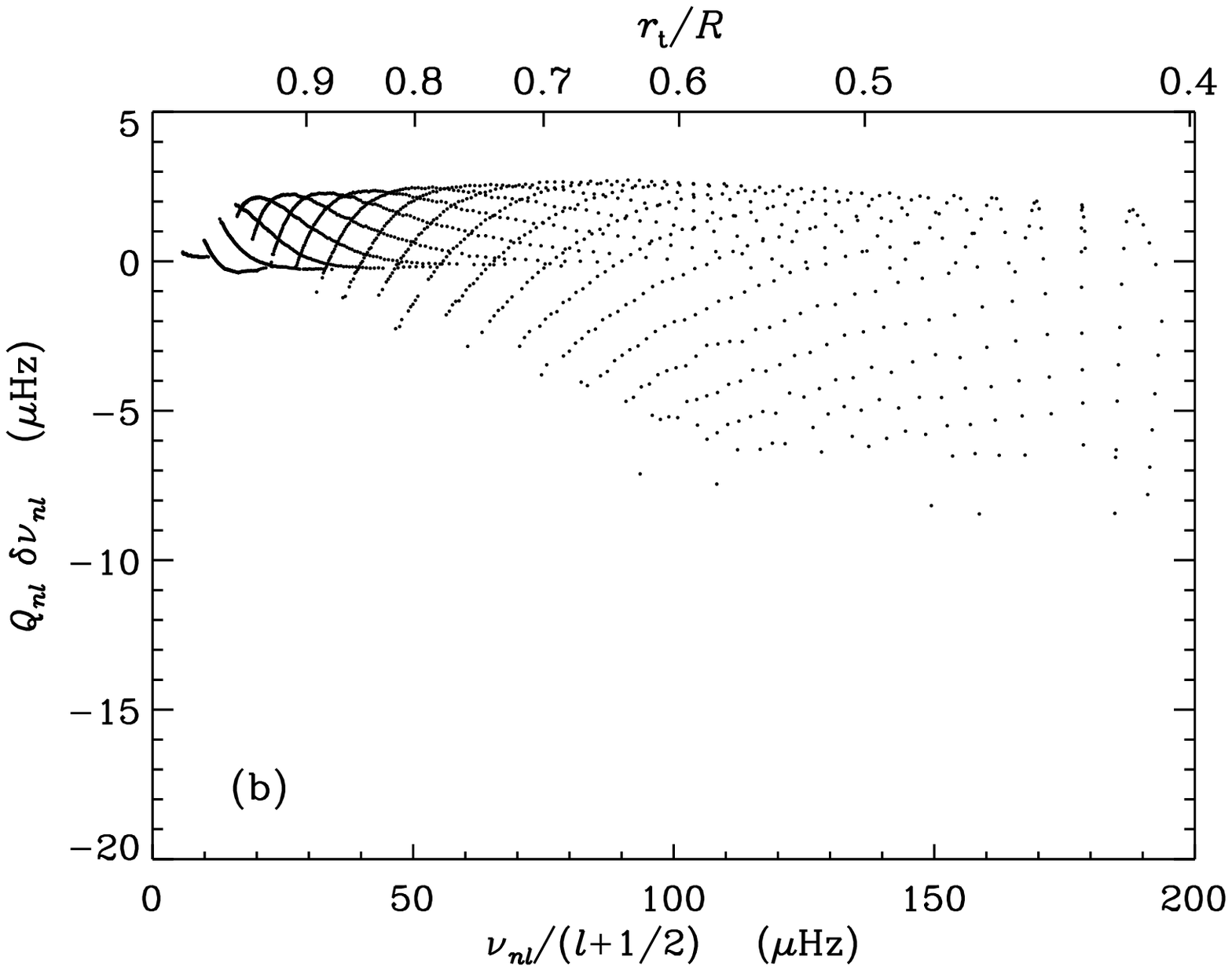}
\caption{Scaled frequency differences $Q_{nl}\delta \nu_{nl}$ between the Sun and models, in the sense (Sun) - (Model), plotted against $\nu/(l+1/2)$. The upper abscissa shows the location of the lower turning point. Panel (a) shows $Q_{nl}\delta \nu_{nl}$ between the Sun and Model S, and panel (b) shows $Q_{nl}\delta \nu_{nl}$ between the Sun and Model NL.}
\label{fig:fdiff2}
\end{figure}

In Fig. \ref{fig:fdiff2}, $Q_{nl}\delta \nu_{nl}$ is plotted against $\nu/L$ (with $L=l+1/2$), and the upper abscissa shows the location of the lower turning point $r_\mathrm{t}$. The near-surface errors are still most obvious here, but we can notice a step at $\nu/L \approx 100~\mu\mathrm{Hz}$ in Fig. \ref{fig:fdiff2}(a); the corresponding $r_\mathrm{t}$ shows the location of the localized sound-speed difference between the Sun and Model S.

The effects of turbulence obtained from 3-D simulations have been parameterized and included in 1-D solar models \citep{Li2002, Robinson2003} in order to explore the effects on solar oscillation frequencies. Alternatively, a patched solar model, in which the SAL of SSM is replaced by simulated SAL, can be used \citep{Rosenthal1999}. Their results showed similar improvements in $Q_{nl}\delta \nu_{nl}$ as shown in Fig. \ref{fig:fdiff}(b). However, \citet{Li2002} found that $F_\mathrm{k}$ was more important than $P_\mathrm{t}$, while $F_\mathrm{k}$ was neglected by \citet{Rosenthal1999}. In our model, $F_\mathrm{k}$ is negligibly small throughout the convection zone and overshooting region, but it works as part of the nonlocal effects of convection. The influences of nonlocal convection on the thermal stratification of the SAL come mainly from $F_\mathrm{c}$ and $P_\mathrm{t}$.

\newpage
\subsection{Detecting the Smoothness of the Transition}
\label{sect:signal}
In MLT models, the abrupt transition at the base of the convection zone leaves a signature on solar acoustic oscillations. \citet{Gough1990} showed that such a rapid variation in solar stratification contributes a characteristic periodic signal $\delta\omega_p$ to the frequencies of oscillation. The amplitude of $\delta\omega_p$ is proportional to the abruptness of the transition. If overshooting is modeled with nonlocal MLT, the amplitude increases because of the discontinuity in the derivative of the sound speed. Thus by calibrating the amplitude of $\delta\omega_p$, the extent of overshooting can be estimated. However, by comparing the observed and model frequencies, it is found that the amplitude of $\delta\omega_p$ in the Sun is comparable with or smaller than that in models without overshooting \citep{MCDT1994, BAN1994, RV1994}, implying that the Sun has a smooth stratification. Therefore a very strong limit (less than one tenth of a pressure scale height) was placed on the extent of the nonlocal MLT overshooting.

\newpage
However, as these authors pointed out, the estimates were based on the assumption that the overshooting zone was adiabatically stratified, which was not justified within the nonlocal MLT. As we have shown here, the lower part of the convection zone is substantially subadiabatic, and the transition to the radiative interior is very smooth. Although the LOZ is quite extended in Model NL, we do not expect $\delta\omega_p$ to be significant.

Following \citet{MCDT1994}, the characteristic periodic signal in solar frequencies caused by abrupt changes is of the form
\begin{equation}
\delta\omega_p=A(\omega)\cos(2\omega\bar{\tau}_d-\bar{\gamma}_d\frac{L^2}{\omega}+2\phi_0),
\label{eq:signal}
\end{equation}
where $L^2=l(l+1)$, $\omega = 2\pi\nu$ is the circular frequency, and $\bar{\tau}_d$, $\bar{\gamma}_d$, and $\phi_0$ are constants. The amplitude $A(\omega)$ consists of two terms:
\begin{equation}
A(\omega)^2=a_1(\frac{\tilde{\omega}}{\omega})^2+a_2(\frac{\tilde{\omega}}{\omega}),
\label{eq:amplitude}
\end{equation}
where $\tilde{\omega}/2\pi= 2500\ \mu\mathrm{Hz}$ is the reference frequency, and $a_1$ and $a_2$ are constants. The parameters ($a_1$, $a_2$, $\bar{\tau}_d$, $\bar{\gamma}_d$, $\phi_0$) are determined by fitting equation (\ref{eq:signal}) to the frequencies after removing a smooth component. The numerical procedures were described in detail in the appendix of \citet{MCDT1994}. A reference value of the amplitude at $\tilde\omega$:
\begin{equation}
A_{2.5}=(a_1^2+a_2^2)^{1/2}
\end{equation}
is used to make comparisons between models and the Sun.

\begin{figure}[htbp]
\centering
\includegraphics[width=0.7\textwidth]{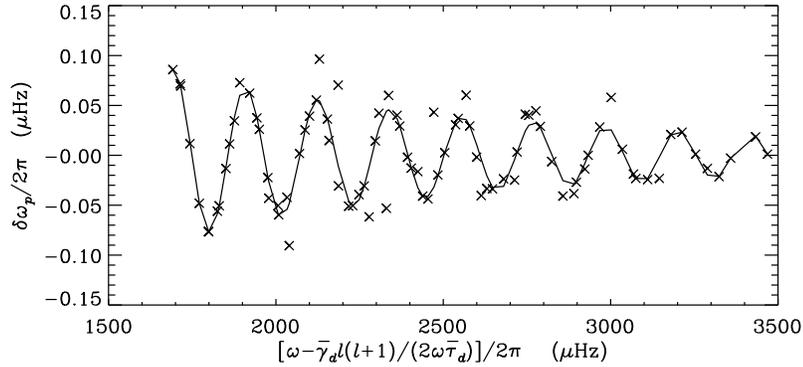}
\caption{Signal of Model NL vs. reduced frequency. The fitted values from equation (\ref{eq:signal}) are shown as the solid line.}
\label{fig:signal}
\end{figure}

We isolate $\delta\omega_p$ from the frequencies of both models and the Sun. We use modes having degree $12 \le l \le 20$ and cyclic frequency between 1700 and 3500 $\mu\mathrm{Hz}$. The signal of Model NL is shown in Fig. \ref{fig:signal} against reduced frequency. The results are listed in Table \ref{tab:fits}. The amplitudes of $\delta\omega_p$ in Model NL and the Sun are comparable, but smaller than that in Model S. Therefore the stratification at the base of the solar convection zone must be smooth, but it does not necessarily lead to the conclusion of limited overshooting.

\begin{table}
\centering
\caption[]{Seismic parameters resulting from the fit of the periodic signal in equation (\ref{eq:signal}) to the frequencies of the Sun and both models.}\label{tab:fits}
\begin{tabular}{@{}lcccc}
\hline
& $A_{2.5}$ & $\bar{\tau}_d$ & $\bar{\gamma}_d$ & $\phi_0$\\
& ($\mu\mathrm{Hz}$) & (s) & ($\mu\mathrm{Hz}$) &\\
\hline
Sun &  0.042 & 2302 & 3.80 & 0.13\\
Model S &  0.068 & 2287 & 8.76 & 0.95\\
Model NL &  0.039 & 2323 & 2.21 & 0.39\\
\hline
\end{tabular}
\end{table}

\label{lastpage}

\newpage
\subsection{Sound-speed Inversion}
\label{sect:inversion}
\begin{figure}[htbp]
\centering
\includegraphics[width=0.7\textwidth]{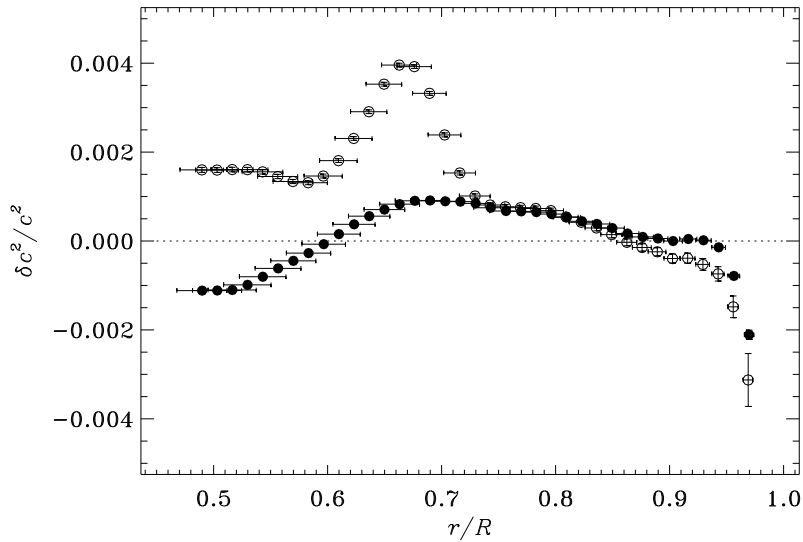}
\caption{Relative differences in squared sound speed, inferred by inversion, between the Sun and models, in the sense (Sun) - (Model). The open symbols are for Model S, and the filled symbols are for Model NL. The vertical error bars correspond to the standard deviations based on the errors in the observed frequencies, while the horizontal bars give a measure of the localization of the solution.}
\label{fig:inversion}
\end{figure}

Helioseismology enables us to deduce the internal structure of the Sun through inverting oscillation frequencies. Such data inversions are based on a linear perturbation analysis of the oscillation equations around a reference model. The differences in the structure between the Sun and the model are related to the differences in the frequencies. Inversion results have shown that the solar structure is remarkably close to the predictions of SSM \citep{Basu1997}, except below the base of the convection zone, where the sound speed of SSM is too low. In Model NL, the sound speed in the LOZ is enhanced as a result of the negative convective flux. Therefore the localized difference in the sound speed between the Sun and SSM is reduced when we use Model NL as the reference model \citep{ZDXCD2012}, as shown in Fig. \ref{fig:inversion}.

\section{Conclusion}
\label{sect:Conclusion}
MLT is an over-simplified phenomenological theory, which does not account for the nonlocal effects of stellar convection. Its simple approximation breaks down at the boundaries of the convection zone.

Our nonlocal RSM constructed with Xiong's closure formulation gives a consistent solution to both the SAL and the LOZ. It shows overshooting at both boundaries of the convection zone, where the convective flux becomes negative. In the SAL, the convective flux decreases due to nonlocal diffusion, and the turbulent pressure is as high as $14\%$ of the gas pressure. These nonlocal effects result in less efficient convection and a steep temperature gradient which helps reduce the frequency differences between the Sun and the model.

Our nonlocal model has an extended overshooting zone below the convective envelope. It shows a smooth transition with substantially subadiabatic stratification. The amplitude of the periodic signal arising from this region is comparable to that of the Sun, and the localized sound-speed difference between the Sun and SSM has been reduced.

Numerical simulations of overshooting can help us gain invaluable insight into the hydrodynamical properties of solar convection, provided solar-like approximations are made. 3-D simulations of the SAL have proven the importance of the turbulent pressure and the existence of overshooting and the steep temperature gradient. Numerical simulations of the LOZ, on the other hand, are not conclusive. Early 2-D and 3-D simulations \citep{RS1993, Hurlburt1994, Saikia2000} found both adiabatic and subadiabatic overshooting, depending on their parameters \citep[cf.][]{Rempel2004}. More resolved simulations of \citet{BCT2002} obtained subadiabatic overshooting with a smooth transition. The subadiabatic stratification has also been confirmed by \citet{CDMRT2011} from the helioseismic point of 
view.

Full 3-D simulations of stellar convection are still computationally unaffordable, therefore RSM may be the only feasible approach that can be implemented in detailed structure and evolution codes without over-simplifying the physical picture \citep{Canuto2007}. However, different RSM formulations tend to give different emphases on turbulent mechanisms, depending on the closure approximations they employ. \citet{ZL2012} showed a very similar picture of overshooting to ours by using the formulation of \citet{LY2007}, in which third-order closure was achieved with a gradient-type scheme; while \citet{MP2002} predicted a very limited overshooting using the formulation of \citet{CD1997, CD1998}, in which the equations were closed at the fourth-order moments. However, high-order closure does not necessarily lead to better results in stellar modeling because it cannot guarantee a good fit of lower-order moments. \citet{Grossman1996} made a detailed comparison between different closure approximations, and found Xiong's closure to be successful because it gave a better representation of the second-order correlations which were most important in constructing stellar models. Nevertheless, more thorough study of the existing closures under stellar conditions is still required. Helio- and asteroseismology and numerical simulations will help us put more constraints on these formulations.

\newpage
Diffusion of helium and heavy elements was not included in the calculation since we used envelope models, but its effects on the solar composition profiles may be limited because of the efficient mixing caused by overshooting. \citet{XD2009} studied the lithium depletion in late-type dwarfs. They found that for cool stars with mass $M \le 1M_\odot$, where $M_\odot$ is the solar mass, the time scale of diffusion is too long, and overshooting is the dominant mechanism of lithium depletion. However, more detailed comparisons need to be made in future evolutionary calculations. Together with the constraints from lithium depletion, we may carry out a more thorough calibration of overshooting.

\begin{acknowledgements}
This work was supported by the Chinese National Natural Science Foundation (CNNSF) under grants 10773029 and 10973015, and the Ministry of Science and Technology of China under grant 2007CB815406. C. Z. received funding from the exchange programme between Chinese Academy of Sciences and the Danish Rectors' Conference (Universities Denmark). Funding for the Stellar Astrophysics Centre is provided by The Danish National Research Foundation (Grant DNRF106). The research is supported by the ASTERISK project (ASTERoseismic Investigations with SONG and Kepler) funded by the European Research Council (Grant agreement no.: 267864).
\end{acknowledgements}

\end{document}